\documentclass[
 pre,
 twocolumn,
 amsmath,
 amssymb,
 aps,
]{revtex4-2}

\usepackage{colortbl}
\usepackage{graphicx} 
\usepackage{booktabs}
\usepackage{caption}
\usepackage{notes2bib}
\usepackage{hyperref}

\usepackage[justification=raggedright]{caption}

\begin{document}

\title{Complexity-stability relationships in competitive disordered dynamical systems}

\author{Onofrio Mazzarisi}
\affiliation{The Abdus Salam International Centre for Theoretical Physics (ICTP), Strada Costiera 11, 34014 Trieste, Italy}
\affiliation{National Institute of Oceanography and Applied Geophysics (OGS), via Beirut 2, 34014 Trieste, Italy}

\author{Matteo Smerlak}
\affiliation{Capital Fund Management, 23 Rue de l'Université, 75007 Paris, France}

\date{\today}

\begin{abstract}
    Robert May famously used random matrix theory to predict that large, complex systems cannot admit stable fixed points. 
    However, this general conclusion is not always supported by empirical observation: from cells to biomes, biological systems are large, complex and, often, stable.
    In this paper, we revisit May's argument in light of recent developments in both ecology and random matrix theory. 
    We focus on competitive systems, and, using a non-linear generalization of the competitive Lotka-Volterra model, we show that there are, in fact, two kinds of complexity-stability relationships in disordered dynamical systems:
    if self-interactions grow faster with density than cross-interactions, complexity is destabilizing; but if cross-interactions grow faster than self-interactions, complexity is stabilizing.
\end{abstract}

\maketitle

\section{Introduction}
Few mathematical arguments have influenced biological thinking like May's prediction that large, complex ecosystems cannot be stable~\cite{May1972}.
It is, on the face of it, a perplexing conclusion.
On the one hand, May's mathematical argument is simple and seemingly model-free, suggesting universal applicability, also beyond biology \cite{Haldane2011, Moran2019}.
On the other hand, it is clear that at least some large, complex systems are stable---else which regularities would biology be studying in the first place? 
In fact, empirical observation suggests the opposite relationship between complexity and stability: species-rich, strongly-coupled communities such as rainforests tend to be stable over time, while sparser ones, for instance arctic communities, often exhibit large fluctuations, extinctions, and invasions~\cite{Hutchinson1959,Odum1959,MacArthur1955}. 
The tension between May's theoretical argument and observation is at the center of the longstanding ``diversity-stability debate" in ecology \cite{McCann2000, Loreau2022,Hatton2024}.

May's argument can be summarized as follows \cite{May1972}.
Consider a system with $N$ populations $x_i$, characterized by an equilibrium point $\mathbf x^*$.
Near that equilibrium $\mathbf x = \mathbf x^* + \delta \mathbf x$, the dynamics of the system is described by linear equations $d(\delta \mathbf x)/dt = A (\delta \mathbf x)$, and the stability of these equations requires that all eigenvalues of $A$ have negative real part.
But if $A$ can be represented as $A = B - I$, where $B$ consists of random, independent interactions (with zero 
mean and variance $\sigma^2$), and $-I$ corresponds to stabilizing self-interactions on some natural timescale,  the circular law of random matrix theory implies that all eigenvalues of $A$ will have negative real part only if $\sigma^2 N < 1$.
This condition places a sharp constraint on both diversity $N$ and interaction strength $\sigma$, often referred to as ``complexity begets instability".
(This argument generalizes to $\langle B_{ij}\rangle \neq 0$, incomplete connectivity, or correlated interactions \cite{allesina2015stability}.)

Many authors have sought to ease the tension between May's prediction and empirical observation by invoking effects not captured by dynamical systems with random coefficients~\cite{McCann2000,Chesson2000,Mougi2012,Rohr2014,Barabas2017,Grilli2017}. 
In this Letter, we use recent results in the physics of disordered systems \cite{Ahmadian2015, Roy2019} to show that May's argument itself is incomplete: in random dynamical systems, stability does not necessarily decrease with dimensionality and interaction strength---the opposite behavior is also possible, without the need for special or additional structure.
For an in-depth discussion of the ecological implications of our result, see our recent paper \cite{Hatton2024}.

\section{Model}
Consider the dynamical system in $N$ variables
\begin{equation}\label{dynamics}
    \dot{x}_i = f(x_i) - \sum_{j}A_{ij}g(x_i)h(x_j) \, .
\end{equation}
Here $f(x_i)$ represents the self-dynamics of a population $i$ (growth and self-regulation), while $g(x_i)$ and $h(x_j)$ capture the cross-interaction of $i$ with other populations. That interaction is weighted by a coefficient $A_{ij}>0$, such that it implies a negative effect of $j$ on the growth of $i$. Non-zero diagonal elements $A_{ii}$ accounts for additional self-interactions with the same functional form as cross-interactions. 
We refer to $f$ as the ``production function'' and $g$ as the ``coupling function''.

This general setting has been used to study universality in network dynamics \cite{Barzel2013} and to construct minimal models of complex dynamics \cite{Barzel2015}, with applications ranging from biological~\cite{Alon2006,Karlebach2008} 
to social~\cite{Pastor-Satorras2001,Hufnagel2004,Dodds2005} systems.
In particular, the classic generalized Lotka-Volterra (GLV) model studied by Bunin and collaborators \cite{bunin2017ecological, biroli2018marginally}, when constrained to competitive interactions, corresponds to $f$, $g$, $h$ all linear. However, it is natural both physically and biologically to consider more general functions, including power laws $f(x)\sim x^\alpha$, $g(x)\sim x^\beta$, $h(x) \sim x^\gamma$.
From a physical perspective, we can imagine populations $x_i$ forming three-dimensional clusters whose growth is limited to their two-dimensional surface, leading to a production function $f(x) \sim x^{2/3}$.
Biologically, at the individual organism level (populations of cells), growth has long been known to scale like $f(x) \sim x^k$ with $k\simeq 3/4$~\cite{Brown2004}, which can be understood in terms of hydrodynamic constraints on vascular and pulmonary networks.
For reasons that are not currently understood, a similar pattern of growth appears to recur at the level of ecological communities~\cite{Hatton2015,Hatton2024}.
In a different direction, it has been recently suggested that predator-prey interactions can be modelled with a square root law, i.e. $g(x) \sim h(x) \sim x^{1/2}$~\cite{Barbier2021,mazzarisi2024}.

\section{Results}
\subsection{Homogeneous interactions}
Under what condition does~\eqref{dynamics} admit a (linearly) stable equilibrium? 
We begin with the simple case where all self-interactions have the same strength $A_{ii} = \mu_s$, and similarly for cross-interactions $A_{ij} = \mu$ ($i\neq j$).
Defining 
\begin{equation}
    \psi(x) \equiv (\mu_s - \mu) -  \left(\frac{f'(x)}{f(x)} - \frac{g'(x)}{g(x)}\right)\frac{f(x)}{g(x)h'(x)},
    \label{eq: psi}
\end{equation}
an elementary calculation shows that stability of the homogeneous equilibrium $x_i^* = x^*$ requires $\psi(x^*)g(x^*)h'(x^*) > 0$, or $\psi(x^*)>0$ if we assume that $g$ and $h$ are positive, increasing functions. This stability condition involves the relative strength of diagonal and off-diagonal interactions ($\mu_s - \mu$), but also on the relative growth rate of the production and coupling functions near the equilibrium ($f'(x^*)/f(x^*) - g'(x^*)/g(x^*)$).

With power laws, the condition $\psi(x^*)>0$ evaluates to 
\begin{equation}
    (\alpha - \beta)(N-1) < \gamma(\mu_s/\mu- 1) - (\alpha - \beta)(\mu_s/\mu),
\end{equation}
leading to three different regimes:
\begin{itemize}
    \item If $\alpha = \beta$, stability requires $\mu_s > \mu$, i.e.
    self-interactions must be stronger than cross-interactions.
    This is the usual conclusion drawn from the Lotka-Volterra model.
    \item If $\alpha > \beta$, stability places an upper bound on $N$: the more complex the system, the less likely to be stable.
    We can call this ``May" behavior.
    \item If $\alpha < \beta$, stability places an lower bound on $N$: the more complex the system, the more likely to be stable.
    This is ``anti-May" behavior.
\end{itemize}

\subsection{Random interactions: stability condition}
We now consider the case of random interactions.
Specifically, we assume that interaction coefficients $A_{ij}$ are drawn independently from a distribution with mean $\mu$ and standard deviation $\sigma$.
We assume the diagonal elements $A_{ii}$ have a mean value $\mu_s$ (possibly different from $\mu$) and standard deviation $\sigma$. 
 
We compute the Jacobian matrix at equilibrium
\begin{align}
    J_{ij}^* & = - A_{ij}g(x_i^*)h'(x_j^*) \qquad \qquad \textrm{for} \ i\neq j \label{eq: jac off-diag}\\
    J_{ii}^* & = f'(x_i^*) - g'(x_i^*)f(x_i^*)/g(x_i^*) - A_{ii}g(x_i^*)h'(x_i^*) \ , \label{eq: jac diag}
\end{align}
where we used $\sum_{j}A_{ij}h(x_j^*)=f(x_i^*)/g(x_i^*)$.
In order to investigate the spectral properties of $J^*$, 
we follow Stone~\cite{Stone2018} in using a recent generalization of the circular law in random matrix theory~\cite{Ahmadian2015}.

In Ref.~\cite{Ahmadian2015}, Ahmadian \emph{et al.} consider matrices of the form $M + LSR$, where $M$,  
$L$ and $R$ are deterministic matrices, and $S$ is a random matrix with i.i.d. coefficients, zero mean and variance $\sigma^2$.
They show that eigenvalues of large matrices of this form are contained in the complex domain $\mathcal{D} = \{\zeta \in \mathbb{C},\, \textrm{Tr}[(\Psi(\zeta) \Psi(\zeta)^\dagger)^{-1}]\geq \sigma^{-2}\}$, where $\Psi(\zeta) = L^{-1}(M-\zeta I)R^{-1}$. If $L$, $R$ and $M$ are all diagonal $N\times N$ matrices, the equation of $\mathcal{D}$ simplifies to 
\begin{equation}
    \sum_{i=1}^N\Big\vert\frac{M_{ii} - \zeta}{L_{ii}R_{ii}}\Big\vert^{-2}\geq \sigma^{-2}.
\label{eq: domain}
\end{equation}

To use this result, we decompose the interaction matrix as $A = \mu\mathbf{1} + (\mu_s-\mu)I + S$,
with $\mathbf{1}$ the matrix with all entries equal to $1$ and $S$ a random matrix as defined above.
Up to the rank-one perturbation $\mu\mathbf{1}$ which does not affect stability properties \cite{Stone2018}, we can write the Jacobian~\eqref{eq: jac off-diag} as $J = M + LSR$ with diagonal matrices
\begin{align}
    M_{ij} = - g(x_i^*)&h'(x_i^*)\psi(x_i^*)\delta_{ij}, \nonumber \\ 
    L_{ij} = g(x_i^*)\delta_{ij}, & \quad R_{ij} = h'(x_i^*)\delta_{ij}, 
\end{align}
where $\delta_{ij}$ is the Kronecker delta and $\psi(x)$ is the function defined in~\eqref{eq: psi}.
\begin{figure}[t!]
    \includegraphics[width=.45\textwidth]{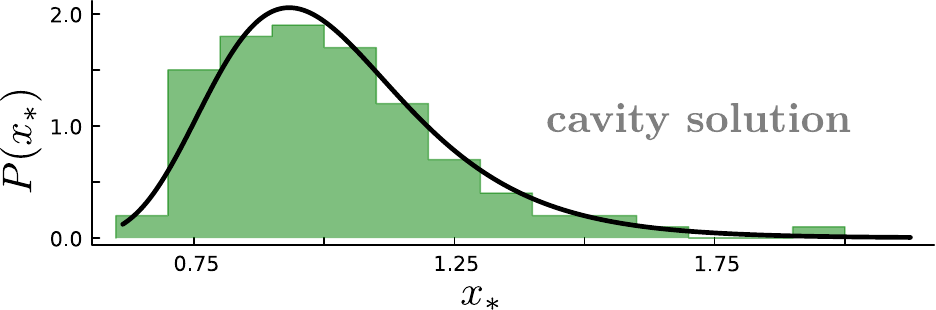}
    \caption{The equilibrium distribution for the case of power-law self- and cross-interactions is accurately reproduced by Eq.~\eqref{eq: dist general}.
    For simulations (in green) we used $\alpha=1$, $\beta=3/2$,
    $\gamma=1$, $N=100$, $\mu=\mu_s=\sigma=10^{-2}$.}
    \label{fig: cavity sol.}
\end{figure}
Stability of the equilibrium requires that $\mathcal{D}$ be entirely contained in the left half-plane (all eigenvalues have negative real part). For this to hold, it is no longer sufficient that $\psi(x_i^*) > 0$ for each $i$: we must also have that $0\notin \mathcal{D}$, hence from~\eqref{eq: domain}:
\begin{equation}
    \sum_{i=1}^N \Big(\frac{\sigma}{\psi(x_i^*)}\Big)^{2}
    < 1.
    \label{eq: random-stability}
\end{equation}
In the GLV model, $\psi(x) = \mu_s - \mu$ (independently of $x$) and we recover the classical condition $\sigma\sqrt{N} < \mu_s - \mu$ \footnote{In fact, the same condition holds when $f$ and $g$ are power laws with the same exponent.}. In general, however, the dependence on equilibrium values $x_i^*$ (which in turn depends on $N$) does not cancel out, and one must gain information about the distribution of equilibrium values $P(x_i^*)$ to assess the stability condition~\eqref{eq: random-stability}.

\subsection{Random interactions: equilibrium distribution}Henceforth we assume $f(x_i)=x_i^{\alpha}$, $g(x_i)=x_i^{\beta}$, $h(x_i)=x_i^{\gamma}$. 
(Any coefficients can be reabsorbed in the statistics of $A$ and by a rescaling of time.)
\begin{figure}[t!]
    \includegraphics[width=.45\textwidth]{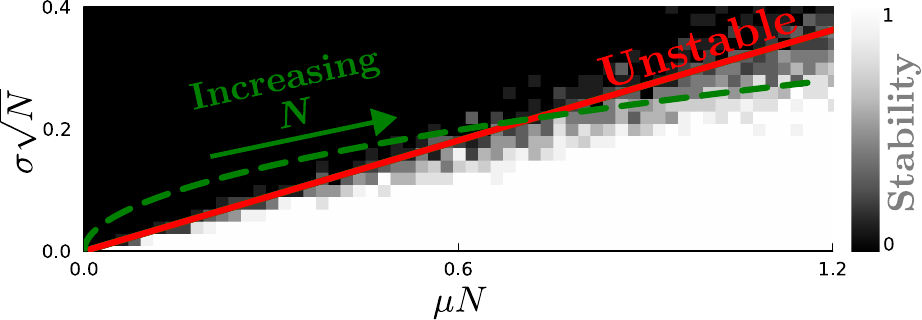}
    \caption{For a system in the ``anti-May" phase, an increase in $N$,
    corresponding to moving along $\sqrt{N}$ in the $(\sigma \sqrt{N},\mu N)$ plane (green dashed line), will be stabilizing rather than destabilizing.
    Here we show results from the numerical resolution of the dynamical system~\eqref{dynamics}. Stability is defined as full stable coexistence. The red line is computed using the cavity solution and the generalized stability condition $N\langle (\sigma/\psi)^2\rangle < 1$.
    Parameters values are $\alpha=1$, $\beta=3/2$,
    $\gamma=1$, $N=50$ and $10$ replicates for the simulations. The green line is plotted for $\mu=0.1$ and $\sigma=0.75$.}
    \label{fig: stability line + sims}
\end{figure}
Following e.g. Ref.~\cite{Roy2019}, we can derive from Eq.~\eqref{dynamics} a cavity solution that describes the ensemble of the stationary solutions of the system by means of a single representative random variable $x_*$
\begin{equation}
    0 = x_*^{\alpha}-A_sx_*^{\beta+\gamma}-x_*^{\beta}\big( \mu N \langle x_*^{\gamma}\rangle + \sigma\sqrt{N} \sqrt{\langle x_*^{2\gamma}\rangle}\xi\big) \, ,
\label{eq: dmft}
\end{equation}
where $A_s$ is a random variable with the statistics of $A_{ii}$ and $\xi$ is a standard normal random variable. 
\begin{figure}[t!]
    \includegraphics[width=.45\textwidth]{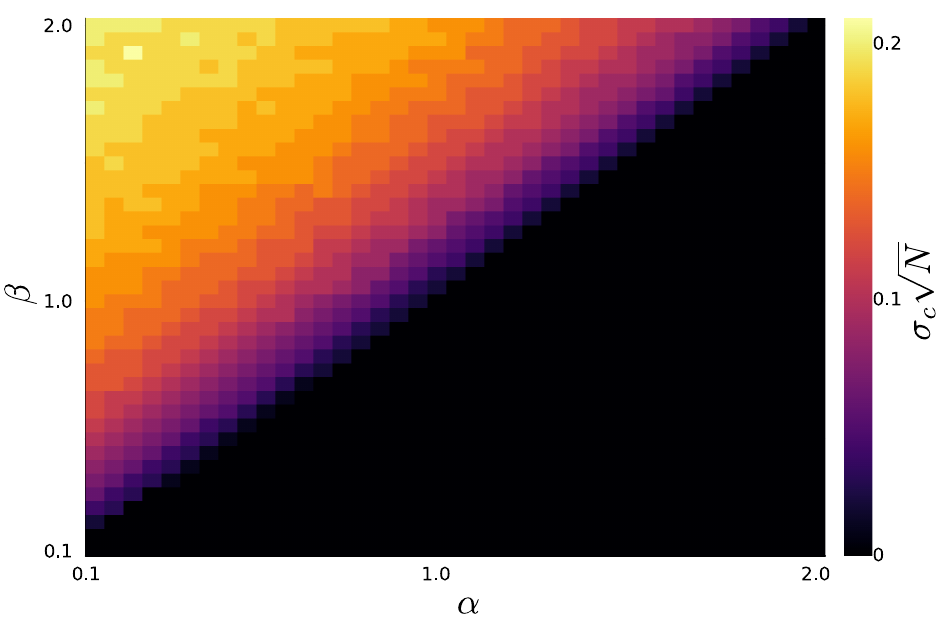}
    \caption{The ``May" and ``anti-May" phases derived in the
    case of uniform interactions ($\sigma = 0$) are robust with respect to random interactions ($\sigma > 0$).
    Here we simulated system~\eqref{dynamics} in the power law case for a fixed value of
    $\mu=\mu_s=10^{-2}$, $\gamma=1$, $N=100$ and $100$ replicates,  
    and we evaluated $\sigma_c$, defined here as
    the value above which full stable coexistence is less probable than 50\%,
    at varying $\alpha$ and $\beta$.
    No amount of heterogeneity is allowed in the bottom right triangle because we have set $\mu=\mu_s$, while an increasing value of $\sigma_c$
    is found as $\beta$ becomes larger than $\alpha$.
    We stress that the existence of a finite $\sigma_c\sqrt{N}$ above which instability is triggered also in the upper triangle,
    does not imply that increasing $N$, at fixed $\mu$ and $\sigma$, will destabilize the system, because $\mu N$ would also change.
    }
    \label{fig: alpha-beta}
\end{figure}
We are interested in the case in which the number of degrees of freedom $N$ is the only parameter changing in the system and the strength and heterogeneity of the interactions are fixed~\cite{kessler2015generalized,fried2016communities}. Compared to the case where $\mu$ and $\sigma$ scale with $N$, our choice reflects a strong-interaction regime; it has been employed, for example, to study ecological scenarios with large differences in species abundance and chaotic turnover~\cite{mallmin2024chaotic}. As we show below, the cavity approximation does a good job at describing the effects of heterogeneity in the interactions as long as $\sigma \sqrt{N}$ is not negligible relative to $\mu N$, at which point we recover the homogeneous case.

The equation above can be solved for $x_*$ for specific values of
$(\alpha-\beta)/\gamma$, or if $A_s=0$.
(In the following we show that neglecting $A_s$, when it is of the order of $\mu$, does not affect noticeably the quality of the approximation.) In this case the stationary solution is given by 
\begin{equation} \label{eq: cavity solution}
    x_* = \left( \mu N \langle x_*^{\gamma}\rangle + \sigma \sqrt{N} \sqrt{\langle x_*^{2\gamma}\rangle}\xi\right)^{1/(\alpha-\beta)} \, .
\end{equation}
The equilibrium distribution $P(x_*)$ can then be obtained as the pushforward of the distribution of $\xi$: 
\begin{equation}\label{eq: dist general}
    P(x_*)=\frac{|\alpha-\beta|x_*^{\alpha-\beta-1}}{\sqrt{2\pi\sigma^2N \langle x_*^{2\gamma}\rangle}}
    \exp{\left\{-\frac{(x_*^{\alpha-\beta}- \mu N \langle x_*^{\gamma}\rangle)^2}{2\sigma^2N\langle x_*^{2\gamma}\rangle}\right\}} \, ,
\end{equation}
defined from $0$  to $\infty$, and where the expectation values must be computed self-consistently. 

Let us focus on the choice $\alpha=1$, $\beta=3/2$ and $\gamma=1$, with $A_s=0$.  
In this case the system is always feasible, and the distribution can be normalized. 
However, the first and second moments formally diverge.
This problem can be overcome by considering that we are dealing with a large, but \emph{finite}, number of degrees of freedom. 
We are therefore interested in predicting the sample mean and variance for a given $N$. Consider the sample mean $\bar{x}_{N}\equiv\sum_{i}^N x_i /N$. We can approximate it with $\bar{x}_N \approx \langle x_* \rangle_{\Lambda} \equiv \int_0^{\Lambda}dx_*x_*P(x_*)$, where the cut-off $\Lambda$ depends on $N$ in such a way that $\int_{\Lambda}^{\infty}dx_*P(x_*)=1/N$, i.e., such that there is statistically less than 1 variable out of $N$ with value above $\Lambda$; see Appendix~\ref{appendix a}.

Following, e.g.,~\cite{Cui2020,Hatton2024} we can compute self-consistently $\Lambda$, $\langle x_*\rangle_{\Lambda}$ and $\langle x_*^2\rangle_{\Lambda}$:
\begin{align}
    \int_{0}^{\Lambda}dx_*P(x_*) &= 1 - \frac{1}{N} , \\    
    \langle x_* \rangle_{\Lambda} &= \int_0^{\Lambda}dx_*x_*P(x_*) , \\
    \langle x_*^2\rangle_{\Lambda} &= \int_0^{\Lambda}dx_*x_*^2P(x_*) .
\end{align}
The resulting distribution is plotted against simulations in Fig.~\ref{fig: cavity sol.}. 
Simulations correspond to $A_s=\mu$, showing that neglecting the $A_s$ term in analytical expressions does not introduce a significant error. Notice that the same ``renormalization'' procedure described above can be extended to other choices of the exponents $\alpha$, $\beta$, and $\gamma$, if needed, to compute $\langle x_*^{\gamma}\rangle$, and $\langle x_*^{2\gamma}\rangle$.

Equipped with the equilibrium distribution $P(x_*)$, we can compute, for given $N$ and $\mu$, the maximal heterogeneity $\sigma_c$ compatible with the linear stability condition $N\langle (\sigma/\psi(x_*))^2\rangle < 1$.

The properties of large random dynamical systems are often portrayed in the $(\sigma \sqrt{N},\mu N)$ plane \cite{bunin2017ecological}.
If we keep the mean $\mu$ and standard deviation $\sigma$ of interactions fixed, an increase in the number of degrees of freedom $N$ moves the system along a square-root trajectory in that plane. In the GLV model, the boundary between the stable and unstable parameter regions is the horizontal line $\sigma_c\sqrt{N} = \mu_s - \mu N / N \approx \mu_s$ \cite{bunin2017ecological}, hence stability is never possible at large $N$. 

In Fig.~\ref{fig: stability line + sims} we plot the stability condition $N\langle (\sigma/\psi(x_*))^2\rangle < 1$ for $\alpha = 1$, $\beta = 3/2$ derived from \eqref{eq: dist general} (solid line) together with results from simulations (shading). Here, because the boundary between the stable and unstable phases is a straight line with positive slope, increasing $N$ will eventually bring the system in the stable region. This behavior corresponds to the ``anti-May" phase defined previously. 

Fig.~\ref{fig: alpha-beta} shows results of simulations for $\sigma_c\sqrt{N}$ in the $(\alpha,\beta)$ plane, at fixed $\mu N$ and $\gamma = 1$. (Numerically, we define $\sigma_c$ as the value of $\sigma$ above which full stable coexistence has a probability lower than $50\%$.) In these simulations, we use $\mu_s = \mu$, which would never be stable in the GLV model. Fig.~\ref{fig: alpha-beta} illustrates the transition between a ``May" phase for $\alpha \geq \beta$ and an ``anti-May" phase for $\alpha < \beta$, showing that our analytical results are robust to heterogeneous interactions.

Finally, we check the robustness of our findings in the scenario in which the exponents are not the same for every variable $x_i$. As a test, we can sample exponents from different Gaussian distributions: $\alpha_i\sim\mathcal N(\alpha,\sigma_e)$, $\beta_i\sim\mathcal N(\beta,\sigma_e)$ and $\gamma_i\sim\mathcal N(\gamma,\sigma_e)$, respectively with mean $\alpha$, $\beta$ and $\gamma$, and, for simplicity, all with the same standard deviation $\sigma_e$. 
Simulations show that our results hold as long as $\sigma_e$ is small enough, in particular, we observe a loss of stability when $\min_i\beta_i>\max_i\alpha_i$ (Appendix~\ref{appendix b}).

\section{Discussion}The relationship between complexity and stability in high-dimensional dynamical systems has been a longstanding puzzle, in ecology and other fields. 
In this Letter, we have shown that the condition $\sigma\sqrt{N}< \mu_s$ does not provide a complete picture of the relationship between complexity and stability. 
In particular, we have seen that the competitive Lotka-Volterra model, often cited in support of May's general prediction, exemplifies a special cancellation in the more general stability condition $N\langle (\sigma/\psi)^2\rangle < 1$.
In models where self-interactions grow slower than cross-interactions~\cite{Hatton2024,samadder2024interconnection}, the opposite behavior is observed: stability becomes \emph{more} likely with increasing diversity $N$.

A modification of our model~\eqref{dynamics} of general ecological relevance describing the case in which per capita growth rates depend on a linear combination of the densities is $\dot{x_i}/x_i=1-F(\sum_jA_{ij}x_j)$, where the function $F$ characterizes the effects of competition.
General models of this kind are used, for example, to study the growth of a clonal population in the presence of a limiting factor~\cite{mazzolini2023universality}.
All the models in this class are destabilized by diversity. (The Jacobian matrix at equilibrium is $J_{ij}^*=-A_{ij}x_i^*F'(\sum_jA_{ij}x_j^*)$, leading to the same simplification of the stability condition~\eqref{eq: random-stability} as in the GLV model.)

Future work can look at the case where two-body interactions are not separable, or include higher order interaction to generalize the results in Ref. \cite{Gibbs2022} based on the GLV model. 

\medskip

The Julia code used for simulations is available at \url{www.github.com/msmerlak/beyond-may}.

\begin{acknowledgments}
We thank Ada Altieri, Matthieu Barbier and Ian Hatton for productive discussions on ecological power laws and the diversity-stability debate and Jacopo Grilli, Vadim Karatayev and Daniel Reuman for feedback on the manuscript.
Funding for this work was provided by the Alexander von Humboldt Foundation in the framework of the Sofja Kovalevskaja Award endowed by the German Federal Ministry of Education and Research, by the Trieste Laboratory on Quantitative Sustainability - TLQS and by NSF BIO OCE grant 2023473.
\end{acknowledgments}

\appendix

\section{CUT-OFF}
\label{appendix a}
In this appendix we consider a case amenable to analytical treatment to exemplify the argument behind the cut-off $\Lambda$ introduced in the main text to deal with diverging moments distributions.

Consider the case of a power law distribution 
\begin{equation}
  P(x)=\frac{x^{-\beta}}{\mathcal{Z}} ,
\end{equation}
defined from 1 to $\infty$ and with 
\begin{equation}
  \mathcal{Z}=\int_1^{\infty}dxx^{-\beta}.  
\end{equation}

\begin{figure}[h!]
  \includegraphics[width=.475\textwidth]{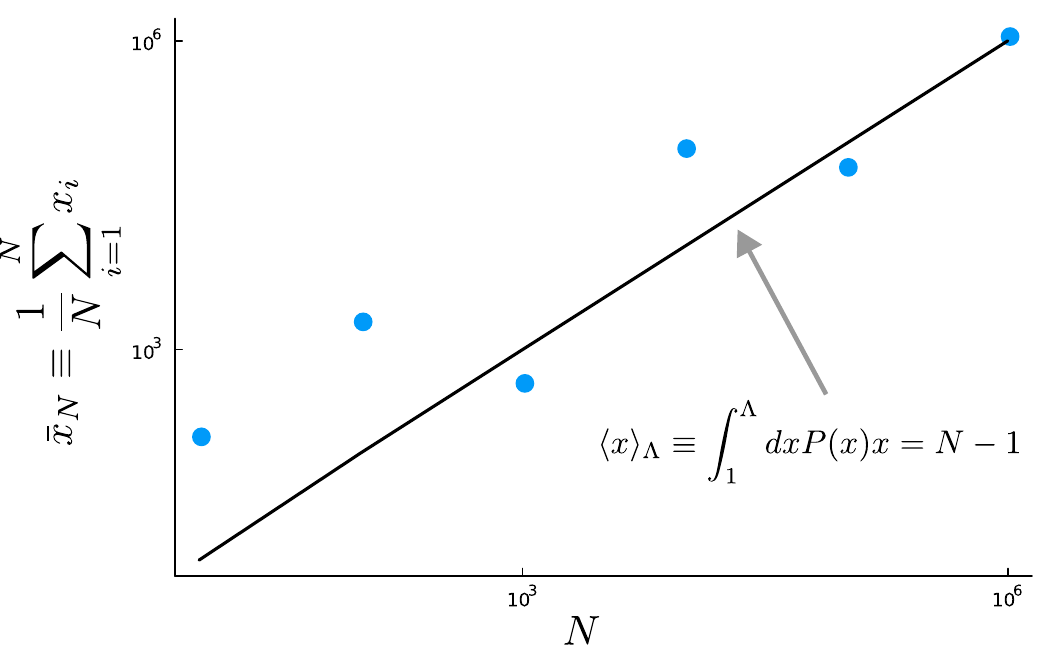}
  \caption{Analytical prediction for the sample mean in Eq.~\eqref{eq: SM-sample-mean} as a function of the number of extractions, $N$, compared with the results from random extractions (dots).
  }
  \label{fig: SM-cut-off}
\end{figure}

Let us choose $\beta=3/2$. Notice that this power law describes the behavior of the distribution we consider in the main text for our example with $\alpha=\gamma=1$ and $\beta=3/2$.
The distribution is normalized with $\mathcal{Z}=2$, but the mean diverges. However, we would like to be able to describe the behavior of the sample mean 
\begin{equation}
  \bar{x}_N\equiv\frac{1}{N}\sum_{i=1}^Nx_i,
\end{equation}

where the $x_i$ are extracted from $P(x)$.
For this purpose, we can define the quantity
\begin{equation}
  \langle x\rangle_{\Lambda}\equiv\int_1^{\Lambda}dxP(x)x,
\end{equation}
with the cut-off $\Lambda$ defined such that $\int_{\Lambda}^{\infty}dxP(x)=1/N$, i.e., such that there is statistically less than 1 variable with value above $\Lambda$ out of $N$ extracted variables. For the case $\beta=3/2$ we have 
\begin{equation}
  \frac{1}{2}\int_{\Lambda}^{\infty}dxx^{-3/2}=\Lambda^{-1/2},
\end{equation}
and therefore $\Lambda=N^2$. We have for the mean
\begin{equation}
  \langle x\rangle_{\Lambda}=\frac{1}{2}\int_1^{N^2}dxx^{-1/2} = N-1.
  \label{eq: SM-sample-mean}
\end{equation}
The result is plotted in Fig.~\ref{fig: SM-cut-off}, alongside the sample mean for extractions of $N=10,$ $10^2,$ $10^3,$ $10^4,$ $10^5,$ $10^6$.

\section{HETEROGENEOUS EXPONENTS}
\label{appendix b}
In this appendix we explore the robustness of our findings in the case in which the exponents characterizing the dynamics of each degree of freedom $x_i$ are not identical for all $i$.

\begin{figure}[h!]
  \includegraphics[width=.475\textwidth]{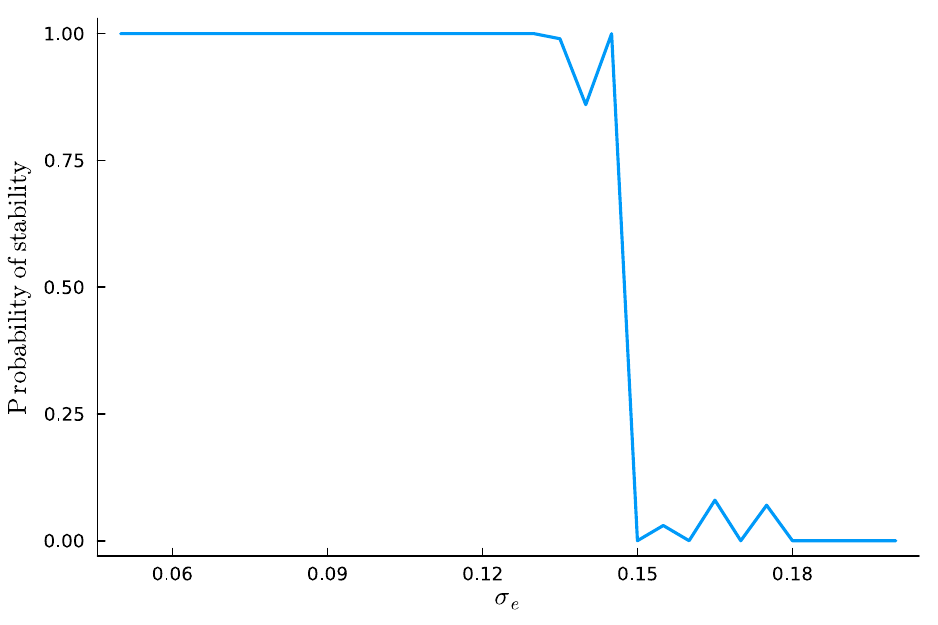}
  \caption{Probability of stability vs. $\sigma_e$ for a system with $S=100$, $\mu=\sigma=0.01$, $\alpha=\gamma=1$ and $\beta=3/2$.
  }
  \label{fig: SM-het-exp}
\end{figure}

Consider the system
\begin{equation}
  \dot{x}_i = x_i^{\alpha_i} - \sum_jA_{ij}x_i^{\beta_i}x_j^{\gamma_i}
\end{equation}
where the exponents $\alpha_i$, $\beta_i$, $\gamma_i$, are extracted from Gaussian distributions: $\alpha_i\sim\mathcal N(\alpha,\sigma_e)$, $\beta_i\sim\mathcal N(\beta,\sigma_e)$ and $\gamma_i\sim\mathcal N(\gamma,\sigma_e)$, respectively with mean $\alpha$, $\beta$ and $\gamma$, and with the same standard deviation $\sigma_e$. The interaction coefficients $A_{ij}$ are drawn independently from a distribution with mean $\mu$ and standard deviation $\sigma$.
Our results are robust when $\sigma_e$ is small enough and we observe a loss of stability when $\min_i\beta_i>\max_i\alpha_i$.
As an example, we show in the plot in Fig.~\ref{fig: SM-het-exp} the probability of stability vs. $\sigma_e$ for a system with $S=100$, $\mu=\sigma=0.01$, $\alpha=\gamma=1$ and $\beta=3/2$.
The probability of stability is obtained as the fraction of stable systems out of $100$ realizations for each value of $\sigma_e$.

\bibliography{beyond-may}
\bibliographystyle{unsrt}

\end{document}